\documentclass{article}
\usepackage{spconf,amsmath,graphicx,cite}
\usepackage{multirow,booktabs}
\usepackage{hyperref}
\usepackage{caption}
\title{PPG-based singing voice conversion with adversarial representation learning}
\name{Zhonghao Li, Benlai Tang, Xiang Yin, Yuan Wan, Ling Xu, Chen Shen, Zejun Ma}
\address{ByteDance AI Lab \\
\tt\small{\{lizhonghao.01,tangbenlai,yinxiang.stephen\}@bytedance.com}
}
\begin{document}
%\ninept
%
\maketitle
\begin{abstract}
Singing voice conversion (SVC) aims to convert the voice of one singer to that of other singers while keeping the singing content and melody. On top of recent voice conversion works, we propose a novel model to steadily convert songs while keeping their naturalness and intonation. We build an end-to-end architecture, taking phonetic posteriorgrams (PPGs) as inputs and generating mel spectrograms. Specifically, we implement two separate encoders: one encodes PPGs as content, and the other compresses mel spectrograms to supply acoustic and musical information. To improve the performance on timbre and melody, an adversarial singer confusion module and a mel-regressive representation learning module are designed for the model. Objective and subjective experiments are conducted on our private Chinese singing corpus. Comparing with the baselines, our methods can significantly improve the conversion performance in terms of naturalness, melody, and voice similarity. Moreover, our PPG-based method is proved to be robust for noisy sources.

\end{abstract}
%

% \vspace{-0.1cm}
\section{Introduction}
\label{sec:intro}
Singing is one of the popular forms of entertainment and self-expression. The goal of singing voice conversion (SVC) is to convert the timbre of a source singer to that of a target singer without changing the content and melody. Compared with conventional speech voice conversion, singing voice conversion requires more considerations about acoustic features. For speech conversion, minor changes of certain features such as pitch and pause, are acceptable. However, for singing conversion, pitch and pause are related to musical characteristics like melody and rhythm, which means they are song-dependent and should be precisely preserved.

The early studies for singing voice conversion generally follow the statistical generation architectures \cite{GMM1,GMM3}, which often use Gaussian mixture model (GMM) with parallel singing data. \cite{SINGAN} updates the conversion framework with Generative Adversarial Network (GAN). However, it still requires the source and the target speakers to sing the same songs during the training phase.

As parallel singing corpus is rare, several works have been conducted to solve this problem. Referring to advanced achievements from voice conversion \cite{autoVC,VC2}, \cite{USVC} builds an autoencoder framework to train the conversion model. The autoencoder model, consisting of a WaveNet \cite{wavenet} encoder to compress acoustic information and a WaveNet decoder to recover waveform with a speaker embedding table, maps the source waveform to itself. With the powerful network architecture, it achieves competitive results with non-parallel data. To enhance the timbre similarity of the converted audio, this work introduces a domain confusion module \cite{confusion} to disentangle singer information from encoder output by an adversarial singer classifier. PitchNet \cite{pitchnet} follows the confusion method and adds an extra pitch confusion module to remove pitch information from the encoder so that it can leverage F0 values to control pitch contour and melody. Moreover, some novel generation frameworks are introduced to the SVC task, such as Gaussian mixture variational autoencoders (GMVAEs) \cite{biencoderDBLSTM} and variational autoencoding Wasserstein GANs (VAW-GANs) \cite{VAWGANVC}. Although the autoencoder-based models can obtain natural singing voices, redundant noise from input data may reduce the quality of the generated sounds.

Another way to address the issue of parallel data limitation is to use phonetic posteriorgrams (PPGs) as the model input \cite{ppgVC}. PPGs represent frame-level linguistic information by probability distributions of phonemes. It deservedly removes acoustic information such as timbre and pitch, while maintaining speaker-independent content and tempo information. For singing voice conversion, \cite{baseline} executes a multi-layer bidirectional LSTM (DBLSTM) network to map PPGs to Mel Cepstrals (MCEPs) in a specific timbre, building a many-to-one singing voice conversion system with WORLD vocoder \cite{world}.
Recently, \cite{crossdomain} upgrades the work to support many-to-many conversion. It uses WaveNet conditioned on various linguistic and acoustic features and presents a non-autoregressive model optimized by several perceptual losses.

\begin{figure*}
% \begin{minipage}[b]{1.0\linewidth}
  \centering
  \centerline{\includegraphics[width=15cm]{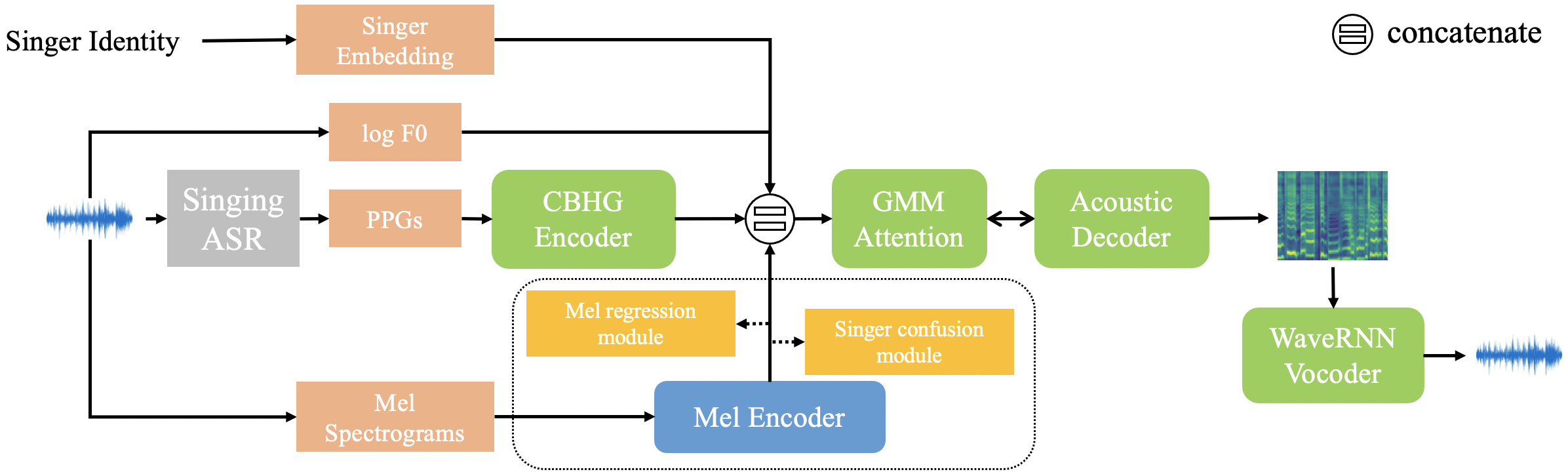}}
% \end{minipage}

\caption{The overall architecture of the proposed model. The framed part is the Mel encoder, and the remaining modules constitute our baseline architecture.}
\label{fig:fig1}

\end{figure*}

In this work, we design a many-to-many SVC model based on the end-to-end framework which is widely used in audio generation tasks \cite{tacotron,tacotron2,DurIANSC,accent}. Benefiting from previous SVC works, our model generally follows a PPG-to-Mel pipeline. An additional reference encoder, named Mel encoder, is implemented to elevate the quality, naturalness, and melody of the conversion outputs. To enhance the timbre similarity, an adversarial singer confusion module \cite{confusion} is applied to disentangle the singer information from the Mel encoder. And then a singer lookup table compensates the singer identity information.
Furthermore, we raise a mel-regressive module to capture to capture acoustic representations from the Mel encoder outputs and singer identity embeddings.
In the experiments, integrating the proposed techniques, our model outperforms the baseline systems in naturalness and timbre-similarity significantly. And an objective evaluation reveals that our model is noise-robust.

\vspace{-0.1cm}
\section{method}
\label{sec:method}

As illustrated in Fig.~\ref{fig:fig1}, the overall structure of the proposed model is a PPG-based end-to-end framework, which takes the source audio as input and outputs the converted audio with the target timbre. The input singing audio is passed through the feature extractors for the input features. Then the conversion model maps the features to mel spectrograms. Finally, A WaveRNN \cite{wavernn} neural vocoder is used to synthesize waveform from the mel spectrograms in real-time and high fidelity.

\vspace{-0.2cm}
\subsection{Proposed Baseline Architecture}
\label{ssec:baseline}

We employ a speaker-independent automatic speech recognition (ASR) model to extract PPGs as linguistic features. Unlike the previous work \cite{baseline}, We train this part with a large scale of singing data. Compared with models trained with speech data, it can improve the conversion results dramatically. To our knowledge, this is the first work using singing data to train the ASR model for the SVC task. The singing ASR (SASR) model is based on DFSMN \cite{dfsmn} with CTC loss \cite{ctc} and includes 30 layers.

For mapping PPGs to mel spectrograms, our fundamental conversion model consists of a linguistic encoder and an acoustic decoder bridged by an attention module. Specifically, we employ a CBHG encoder \cite{tacotron} to encode the frame-level PPG features to the linguistic representation. Besides, the singer identity embedding is selected from the singer lookup table. And the logarithmic F0 sequence is extracted from the source audio. Then the singer embeddings and the F0 sequence are concatenated to the encoder outputs. The acoustic decoder follows the design in \cite{tacotron2}. GMM attention mechanism \cite{GMMattention} is used for its capacity of generating very long utterances \cite{GMMattention2}.

Mel spectrograms regression loss and stop token prediction loss are used to optimize the conversion model as \cite{tacotron2}. The loss to be minimized can be described as
\begin{equation}
    L_{dec} = MSE({\bf Y}_{Dout}, {\bf Y}_{target}) + CE({\bf t}_{out}, {\bf t}_{target}) \label{baseloss}
\end{equation}
where ${\bf Y}$ represents mel spectrograms, and ${\bf t}$ represents stop tokens. ${\bf Y}_{Dout}$ is the output of the acoustic decoder. Mean Square Error (MSE) measures the difference between generated and target mel spectrograms. And binary Cross Entropy (CE) is used for the stop token prediction.

\vspace{-0.2cm}
\subsection{Mel Encoder}
\label{ssec:refenc}

PPG features ideally disentangle undesired timbre information from the source voice. However, it is difficult to reconstruct the style from the singing source (e.g. intonation, melody, emotion, etc.) that are not covered abundantly by the present features. To preserve these musical characteristics from the source audio, we use an additional encoder to extract information from the source mel spectrograms, called Mel encoder.

\begin{figure}
    % \begin{minipage}[b]{1.0\linewidth}
    \centering
    \centerline{\includegraphics[width=8.5cm]{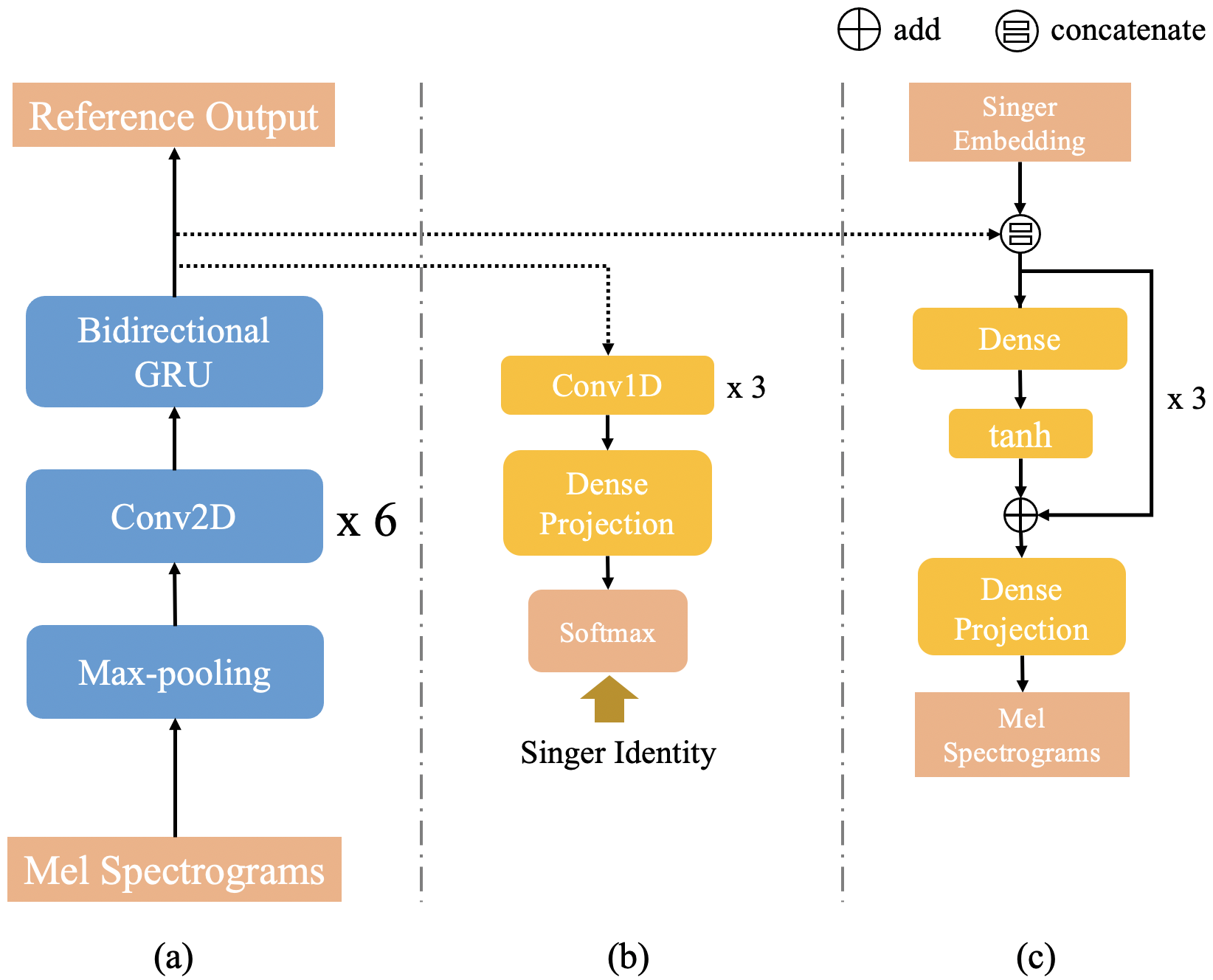}}
    % \end{minipage}
    
    \caption{Our proposed adversarial representation learning encoder. (a) Mel encoder architecture. (b) The singer confusion module. (c) The mel-regressive representation learning module. (b) and (c) are not used in the inference phase.}
    \label{fig:fig2}
\end{figure}

The structure of Mel encoder is shown in Fig.~\ref{fig:fig2}(a). First, the mel spectrogram extracted from the source audio is fed into a max-pooling layer, followed by 6 2-D convolution layers and a bidirectional GRU network. The outputs are concatenated with the encoder outputs as described in Section~\ref{ssec:baseline}.

It is worth noting that the dimension of the Mel encoder output is set to be minimal, to suppress the effects of the timbre and sound noise of the source voice. As evaluated in \cite{accent}, we found 4 units performed best in balancing timbre, sound quality, and musical characters of the converted audio.

\subsection{Singer Confusion Module}
\label{ssec:gan}

To strengthen the timbre similarity of the output, a singer confusion module is introduced to our model. During the training phase, the encoded embeddings from Mel encoder are fed into a singer-identity classifier. The classifier is illustrated in Fig~\ref{fig:fig2}(b). The input embeddings are passed through three 1-D convolutional layers followed by a dense projection and transferred to represent probability distributions in the size of singer identity classes.

With ${\bf c}$ being the probability distribution of the singer identity, ${\bf c}_{target}$ is a one-hot vector. For an embedding sequence with $N$ frames, ${\bf c}_{out}^j$ is the classifier output, where $j=1,2,...,N$. The cross entropy values between ${\bf c}_{out}^j$ and ${\bf c}_{target}$ are averaged for the classification loss
\begin{equation}
    L_D = \sum_{j=1}^N{CE({\bf c}_{out}^j, {\bf c}_{target})} / N
\end{equation}
Training with the confusion module is a cycle of two steps. First, the classification network is trained to minimize $L_D$. Second, the conversion pathway except the classifier part is trained with the loss
\begin{equation}
    L_G = L_{dec} - \lambda L_D
\end{equation}
where $\lambda$ is a weight factor. The entire model forms an adversarial framework. The baseline conversion model with Mel encoder is the generator. And the singer-identity classifier becomes a discriminator to distinguish singers with Mel encoder outputs. We also attempt to employ this module to the CBHG encoder, but it leads to instability and even collapse during the training phase.

\subsection{Mel-Regressive Representation learning Module}
\label{ssec:specpred}

Although the singer confusion technique improves timbre similarity and stability, the naturalness of the converted audio is evidently declined.
It indicates the confusion module decreases articulatory and musical representation of the Mel encoder, besides disentangling singer identities. To handle the issue, we propose a novel representation learning decoder.
% To handle the issue, we enhance the ability of the Mel encoder by proposing a representation learning module.

As shown in Fig.~\ref{fig:fig2}(c), the additional part is a mel-regressive network, which consists of 3 residual activated DNN layers and a projection layer. 
In the training phase, output embeddings of the Mel encoder are concatenated with the singer embedding to compensate identity information and then passed through the regressive network to generate the mel spectrograms ${\bf Y}_{Mout}$.
The regression loss is calculated by the MSE function.
\begin{equation}
    L_{melEnc} = MSE({\bf Y}_{Mout}, {\bf Y}_{target})
\end{equation}
By training Mel encoder with the regression loss, the Mel encoder outputs are expected to preserve acoustic and musical information for reconstructing mel spectrograms except for singer identity information.

Integrating all proposed modules, the overall conversion model is trained with an augmented generation loss
\begin{equation}
    L_G = L_{dec} + \gamma L_{melEnc} - \lambda L_D
\end{equation}
and $\gamma$ is the weight factor of the mel-regressive module.

\vspace{-0.2cm}
\section{Experiments}
\label{sec:exp}

\subsection{Experimental Setup}
\label{ssec:setup}

Our experiments are implemented with an internal Chinese mandarin singing corpus. 
The data set consists of totally 32.7-hour audio data from 9 female singers and 7 male singers. And each singer has 1000 utterances for training on average and 10 for validation. For evaluation, we choose a female singer and a male singer as the target timbres. The test set consists of 40 segments from 20 singers out of the training set\footnote{Samples can be found in \href{https://lzh1.github.io/singVC}{https://lzh1.github.io/singVC}}. All songs are sampled to 24kHz.

To obtain input features, the SASR model is trained with about 20k hours of singing data, generating 1467-dimensional PPGs as outputs. And F0 values are extracted by REAPER\footnote{\href{https://github.com/google/REAPER}{https://github.com/google/REAPER}}. Configurations for extracting mel spectrograms and training WaveRNN model configuration follow \cite{accent}. The conversion model is trained for 200k steps with a batch size of 32. We use Adam optimizer \cite{adam} and halve the learning rate per 25k steps. For weight factors in Section~\ref{sec:method}, we set $\gamma = 1.0, \lambda = 0.1$.

\begin{table}[]
    \small
    \newcommand{\tabincell}[2]{\begin{tabular}{@{}#1@{}}#2\end{tabular}}
    \caption{MOS evaluation scores of the baseline models and our proposed model. GT means ground truth.}
    \label{tab:main}

    \centering
    \begin{tabular}{l l|c c|c}
        % \hline
        \specialrule{0.1em}{1pt}{2pt}
        {} & {\bf System} & {\bf Naturalness} & {\bf Similarity} & {\bf NCC} \\
        % \hline
        \specialrule{0.1em}{2pt}{2pt}
        Source & GT & 4.43$\pm$0.11 & 2.97$\pm$0.19 & - \\
        % \hline
        \specialrule{0.05em}{2pt}{2pt}
        \multirow{5}{*}{\tabincell{l}{Female \\ target}} & GT & 4.41$\pm$0.07 & 4.12$\pm$0.09 & - \\
        ~ & BASE1 & 2.79$\pm$0.18 & 2.47$\pm$0.19 & 0.927 \\
        ~ & BASE2 & 2.85$\pm$0.13 & 2.43$\pm$0.16 & 0.930 \\
        ~ & BASE3 & 2.58$\pm$0.08 & 3.55$\pm$0.10 & 0.850 \\
        ~ & Proposed & 3.75$\pm$0.18 & 3.57$\pm$0.19 & 0.902 \\
        % \hline
        \specialrule{0.05em}{2pt}{2pt}
        \multirow{5}{*}{\tabincell{l}{Male \\ target}} & GT & 4.49$\pm$0.08 & 4.24$\pm$0.09 & - \\
        ~ & BASE1 & 3.10$\pm$0.12 & 2.69$\pm$0.11 & 0.924 \\
        ~ & BASE2 & 3.21$\pm$0.13 & 2.75$\pm$0.11 & 0.926 \\
        ~ & BASE3 & 2.35$\pm$0.09 & 2.96$\pm$0.14 & 0.870 \\
        ~ & Proposed & 3.64$\pm$0.15 & 3.42$\pm$0.09 & 0.889 \\
        % \hline
        \specialrule{0.1em}{2pt}{0pt}
    \end{tabular}

    \vspace{-0.1cm}
\end{table}

\subsection{Evaluations}

\noindent {\bf Subjective and Objective Evaluations} We compare our model with three baseline systems: (1) BASE1, the DBLSTM model \cite{baseline} which uses PPGs for the SVC task for the first time. In this work, we use our SASR model for PPG extraction instead. (2) BASE2, augmented from BASE1 to support multi-singer corpora by adding a singer lookup table to promote the transformation. (3) BASE3, our proposed baseline system introduced in Section~\ref{ssec:baseline}.

For subjective evaluations, all of the converted samples from each system are scored individually by 18 music professionals. Two metrics are conducted to measure the models: (1) Mean Opinion Score (MOS) of the naturalness, used for judging an integrated assessment of intonation, rhythm, melody, clarity, and expression. (2) MOS of the timbre similarity between the converted samples and the target singing voices. Both metrics are scaled between 1-5. Moreover, we use the normalized cross-correlation (NCC) as an objective evaluation to measure the pitch match between the prediction and the ground truth.

Conducted scores are illustrated in Tab.~\ref{tab:main}. Comparing BASE2 with BASE1, the former performs better both in naturalness and similarity, explaining our proposal to use multi-singer training data. The DBLSTM models only convert MCEP features and use the pitch values from the source audio, resulting in high accuracy in intonation and melody.
BASE3 gets the lowest NCC scores and naturalness MOS compared with BASE1 and BASE2, showing its shortage in maintaining pitch and voice quality.
However, in terms of similarity scores, BASE3 outperforms the former methods substantially, as it employs a more complicated singer-dependent decoder than the DBLSTM network.
Moreover, our entire framework gains the highest scores in the two subjective metrics, and also obtains competitive scores in NCC. The results show that our framework with additional modules supplying more musical information from mel spectrograms, is promoted in naturalness and pitch precision.

\begin{table}[]
    \small
    \centering
    \caption{Ablation Tests. ME means Mel encoder, SC denotes the singer confusion module and MS denotes the mel-regressive module. The additions are accumulations.}
    \label{tab:ablation}

    \begin{tabular}{l l|c c|c}
        % \hline
        \specialrule{0.1em}{0pt}{2pt}
        {\bf Target} & {\bf System} & {\bf Naturalness} & {\bf Similarity} & {\bf NCC}  \\
        \specialrule{0.1em}{2pt}{2pt}
        % \hline
        \multirow{5}{*}{Female} & BASE3 & 2.58$\pm$0.08 & 3.55$\pm$0.10 & 0.850 \\
        ~ & +ME & 2.81$\pm$0.33 & 2.61$\pm$0.27 & 0.865 \\
        ~ & +SC & 2.52$\pm$0.10 & 3.52$\pm$0.13 & 0.879 \\
        ~ & +MS & 3.75$\pm$0.18 & 3.57$\pm$0.19 & 0.901 \\
        % \hline
        \specialrule{0.05em}{2pt}{2pt}
        \multirow{5}{*}{Male} & BASE3 & 2.35$\pm$0.09 & 2.96$\pm$0.14 & 0.870 \\
        ~ & +ME & 2.93$\pm$0.22 & 2.74$\pm$0.13 & 0.895 \\
        ~ & +SC & 2.41$\pm$0.10 & 3.04$\pm$0.13 & 0.872 \\
        ~ & +MS & 3.64$\pm$0.15 & 3.42$\pm$0.09 & 0.889 \\
        \specialrule{0.1em}{2pt}{3pt}
    \end{tabular}

    \vspace{-0.1cm}
\end{table}

\noindent {\bf Ablation} Ablations are executed to present contributions of the proposed modules. Tab.~\ref{tab:ablation} summarises the evaluation results. As shown by the scores, Mel encoder improves the quality of the converted samples but decreases timbre similarity. By adding the adversarial singer confusion module, the timbre similarities of the converted samples acquire great progress. However, both the NCC score and the naturalness MOS fall because of the disentangling process on Mel encoder. The Mel-regressive module eliminates the adverse effect of the confusion module and further improves the performance extremely and comprehensively.

\begin{table}[]
    \small
    \caption{SNR values of source voices and converted voices.}
    \label{tab:snr}

    \centering
    \begin{tabular}{c|c|c}
        % \hline
        \specialrule{0.1em}{-1pt}{2pt}
        {\bf Source} & {\bf Female target} & {\bf Male target} \\
        \specialrule{0.1em}{2pt}{2pt}
        % \hline
        25.35 & 30.41 & 33.56 \\
        15.30 & 24.41 & 27.38 \\
        8.18 & 22.42 & 19.64 \\
        % \hline
        \specialrule{0.1em}{2pt}{0pt}
    \end{tabular}
    \vspace{-0.2cm}
\end{table}

\noindent {\bf Noise Robustness} To show the robustness of our method in handling noise, our proposed model is used to convert source audio adding various levels of white noise. Signal-to-noise ratio (SNR) is calculated to measure the clarity of the source and converted samples. Results in Tab.~\ref{tab:snr} indicate SNR of the converted samples falls slightly when the added noise increases.

\vspace{-0.2cm}
\section{Conclusion and Future Work}
\label{sec:conc}

We devote a novel end-to-end model for singing voice conversion in this work. PPGs extracted from a singer independent SASR model are used as input features thus the model can achieve many-to-many singing conversion. The proposed model contains an additional encoder to obtain acoustic and musical information from mel spectrograms, along with a singer confusion module and a mel-regressive representation learning module. Experiments show that our proposed model outperforms the baseline models significantly, generating natural and pitch-accurate singing voices in the target timbre. We also confirm that our proposed system can make conversion robustly. In future work, we will continue to improve the quality of converted audio and attempt to use less data for better performance.

% \vfill\pagebreak

\bibliographystyle{IEEEbib}
\bibliography{refs}

\end{document}